\documentclass[journal, draftcls, one column, 12pt]{IEEEtran}

\usepackage{amsmath}
\usepackage{graphicx}
\usepackage{amssymb}
\usepackage{cases}
\usepackage{color}
\usepackage{cite}
\usepackage[ruled,vlined,lined,ruled,linesnumbered]{algorithm2e}
\usepackage{relsize}
\usepackage[nodisplayskipstretch]{setspace}
\usepackage[top=1in, bottom=1in, left=0.7in, right=0.7in]{geometry}

\begin{document}
\title{Placement Optimization of Energy and Information Access Points in Wireless Powered Communication Networks}
\author{Suzhi~Bi, ~\IEEEmembership{Member,~IEEE} and Rui Zhang, ~\IEEEmembership{Senior Member,~IEEE} \\
                \thanks{This work will be presented in part at the IEEE Global Communications Conference (GLOBECOM), San Diego, CA, USA, Dec. 6-10, 2015. This work was supported in part by the National Natural Science Foundation of China (Project number 61501303).}
        \thanks{S.~Bi is with the College of Information Engineering, Shenzhen University, Shenzhen, Guangdong, China 518060 (e-mail: bsz@szu.edu.cn).}
        \thanks{R.~Zhang is with the Department of Electrical and Computer Engineering, National University of Singapore, Singapore 117583, and also with the Institute for Infocomm Research, A$^*$STAR, Singapore 138632 (e-mail:elezhang@nus.edu.sg).} }

\maketitle

\vspace{-25pt}

\begin{abstract}
The applications of wireless power transfer technology to wireless communications can help build a wireless powered communication network (WPCN) with more reliable and sustainable power supply compared to the conventional battery-powered network. However, due to the fundamental differences in wireless information and power transmissions, many important aspects of conventional battery-powered wireless communication networks need to be redesigned for efficient operations of WPCNs. In this paper, we study the placement optimization of energy and information access points in WPCNs, where the wireless devices (WDs) harvest the radio frequency energy transferred by dedicated energy nodes (ENs) in the downlink, and use the harvested energy to transmit data to information access points (APs) in the uplink. In particular, we are interested in minimizing the network deployment cost with minimum number of ENs and APs by optimizing their locations, while satisfying the energy harvesting and communication performance requirements of the WDs. Specifically, we first study the minimum-cost placement problem when the ENs and APs are separately located, where an alternating optimization method is proposed to jointly optimize the locations of ENs and APs. Then, we study the placement optimization when each pair of EN and AP are co-located and integrated as a hybrid access point, and propose an efficient algorithm to solve this problem. Simulation results show that the proposed methods can effectively reduce the network deployment cost and yet guarantee the given performance requirements, which is a key consideration in the future applications of WPCNs.
\end{abstract}

\begin{IEEEkeywords}
Wireless power transfer, wireless powered communication networks, energy harvesting, network planning, node placement optimization.
\end{IEEEkeywords}

\section{Introduction}
Modern wireless communication systems, e.g., cellular networks and wireless sensor networks (WSNs), are featured by larger bandwidth, higher data rate and lower communication delays. The improvement on communication quality and the increased data processing complexity have imposed higher requirement on the quality of power supply to wireless devices (WDs). Conventionally, WDs are powered by batteries, which have to be replaced/recharged manually once the energy is depleted. Alternatively, the recent advance of radio frequency (RF) enabled wireless power transfer (WPT) provides an attractive solution to power WDs over the air \cite{2014:Bi,2015:Lu}. By leveraging the far-field radiative properties of microwave, WDs can harvest energy remotely from the RF signals radiated by the dedicated energy nodes (ENs) \cite{2013:Zhou}. Compared to the conventional battery-powered methods, WPT can save the cost due to manual battery replacement/recharging in many applications, and also improve the network performance by reducing energy outages of WDs. Currently, tens of microwatts ($\mu$W) RF power can be effectively transferred to a distance of more than $10$ meters.\footnote{Based on the product specifications on the website of Powercast Co. (http://www.powercastco.com), with TX91501-3W power transmitter and P2110 Powerharvester receiver, the harvest RF power at a distance of $10$ meters is about $40\ \mu$W.} The energy is sufficient to power the activities of many low-power communication devices, such as sensors and RF identification (RFID) tags. In the future, we expect more practical applications of RF-enabled WPT to wireless communications thanks to the rapid developments of many performance enhancing technologies, such as energy beamforming with multiple antennas \cite{2013:Zhang} and more efficient energy harvesting circuit designs \cite{2010:Georgiadis}.

In a wireless powered communication network (WPCN), the operations of WDs, including data transmissions, are fully/partially powered by means of RF-enabled WPT \cite{2014:Ju1,2014:Liu2,2014:Ju2,2014:Huang1,2014:Lee1,2015:Che,2015:Nasir,2014:Krikidis,2015:Chen}. A TDMA (time division multiple access) based protocol for WPCN is first proposed in \cite{2014:Ju1}, where the WDs harvest RF energy broadcasted from a hybrid access point (HAP) in the first time slot, and then use the harvested energy to transmit data back to the HAP in the second time slot. Later, \cite{2014:Liu2} extends the single-antenna HAP in \cite{2014:Ju1} to a multi-antenna HAP that enables more efficient energy transmission via energy beamforming as well as more spectrally efficient SDMA (space division multiple access) based information transmission as compared to TDMA. To further improve the spectral efficiency, \cite{2014:Ju2} considers using full-duplex HAP in WPCNs, where a HAP can transmit energy and receive user data simultaneously via advanced self-interference cancelation techniques. Intuitively, using a HAP (or co-located EN and information AP), instead of two separated EN and information access point (AP), to provide information and energy access is an economic way to save deployment cost, and the energy and information transmissions in the network can also be more efficiently coordinated by the HAP. However, using HAP has an inherent drawback that it may lead to a severe ``doubly-near-far" problem due to distance-dependent power loss \cite{2014:Ju1}. That is, the far-away users quickly deplete their batteries because they harvest less energy in the downlink (DL) but consume more energy in the uplink (UL) for information transmission. To tackle this problem, separately located ENs and APs are considered to more flexibly balance the energy and information transmissions in WPCNs \cite{2014:Huang1,2014:Lee1,2015:Che}. In this paper, we consider the method using either co-located or separate EN and information AP to build a WPCN.

Most of the existing studies on WPCNs focus on optimizing real-time resource allocation, e.g., transmit signal power, waveforms and time slot lengths, based on instantaneous channel state information (CSI, e.g., \cite{2014:Ju1,2014:Liu2,2014:Ju2}). In this paper, we are interested in the long-term network performance optimization based mainly on the average channel gains. It is worth mentioning that network optimizations in the two different time scales are complementary to each other in practice. That is, we use long-term performance optimization methods for the initial stage of network planning and deployment, while using short-term optimization methods for real-time network operations after the deployment. Many current works on WPCNs use stochastic models to study the long-term performance because of the analytical tractability, especially when the WDs are mobile in location. For instance, \cite{2014:Huang1} applies a stochastic geometry model in a cellular network to derive the expression of transmission outage probability of WDs as a function of the densities of ENs and information APs. Similar stochastic geometry technique is also applied to WPT-enabled cognitive radio network in \cite{2014:Lee1} to optimize the transmit power and node density for maximum primary/secondary network throughput. However, in many application scenarios, the locations of the WDs are fixed, e.g., a sensor network with sensor (WD) locations predetermined by the sensed objects, or an IoT (internet-of-things) network with static WDs. In this case, a practical problem that directly relates to the long-term performance of WPCNs, e.g., sensor's operating lifetime, is to determine the optimal locations of the ENs and APs. Nonetheless, this important node placement problem in WPCNs is still lacking of concrete studies.

In conventional battery-powered wireless communication networks, node placement problem concerns the optimal locations of information APs only, which has been well investigated especially for wireless sensor networks using various geometric, graphical and optimization algorithms (see e.g., \cite{2005:Pan,2004:Bogdanov,2007:Akkaya,2003:Gandham,2010:Lin}). However, there exist major differences between the node placement problems in battery-powered and WPT-enabled wireless communication networks. On one hand, a common design objective in battery-powered wireless networks is to minimize the highest transmit power consumption among the WDs to satisfy their individual transmission requirements. However, such energy-conservation oriented design is not necessarily optimal for WPCNs, because high power consumption of any WD can now be replenished by means of WPT via deploying an EN close to the WD. On the other hand, unlike information transmission, WPT will not induce harmful co-channel interference to unintended receivers, but instead can boost their energy harvesting performance. These evident differences indicate that the node placement problem in battery-powered wireless communication networks should be revisited for WPCNs, to fully capture the advantages of WPT.

In this paper, we study the node placement optimization problem in WPCNs, which aims to minimize the deployment cost on ENs and APs given that the energy harvesting and communication performances of all the WDs are satisfied. Our contributions are detailed below.
\begin{enumerate}
  \item We formulate the optimal node placement problem in WPCNs using either separated or co-located EN and AP. To simplify the analysis, we then transform the minimum-cost deployment problem into its equivalent form that optimizes the locations of fixed number of ENs and APs;
  \item The node placement optimization using separated EN and AP is highly non-convex and hard to solve. To tackle the non-convexity of the problem, we first propose an efficient cluster-based greedy algorithm to optimize the locations of ENs given fixed AP locations. Then, a trial-and-error based algorithm is proposed to optimize the locations of APs given fixed ENs locations. Based on the obtained results, we further propose an effective alternating method that jointly optimizes the EN and AP placements;
  \item For the node placement optimization using co-located EN and AP (or HAP), we extend the greedy EN placement method under fixed APs to solving the HAP placement optimization, which is achieved by incorporating additional considerations of dynamic WD-HAP associations during HAP placement. Specifically, a trial-and-error method is used to solve the WD-HAP association problem, which eventually leads to an efficient greedy HAP placement algorithm.
\end{enumerate}
Due to the non-convexity of the node placement problems in WPCNs, all the proposed algorithms are driven by the consideration of their applicabilities to large-size WPCNs, e.g., consisting of hundreds of WDs and EN/AP nodes. Specifically, we show that the proposed algorithms for either separated or co-located EN and AP placement are convergent and of low computational complexity. Besides, simulations validate our analysis and show that the proposed methods can effectively reduce the network deployment cost to guarantee the given performance requirements. The proposed algorithms may find their wide application in the future deployment of WPCNs, such as wireless sensor networks and IoT networks.

The rest of the paper is organized as follows. In Section II, we first introduce the system models of WPCN where the ENs and APs are either separated or co-located. Then, we formulate the optimal node placement problems for the two cases in Section III, and propose efficient algorithms to solve the problems in Sections IV and V, respectively. In Section VI, simulations are performed to evaluate the performance of the proposed node placement methods. Finally, we conclude the paper and discuss future working directions in Section VII.

\section{System Model}

\subsection{Separated ENs and APs}
\begin{figure}
\centering
  \begin{center}
    \includegraphics[width=0.6\textwidth]{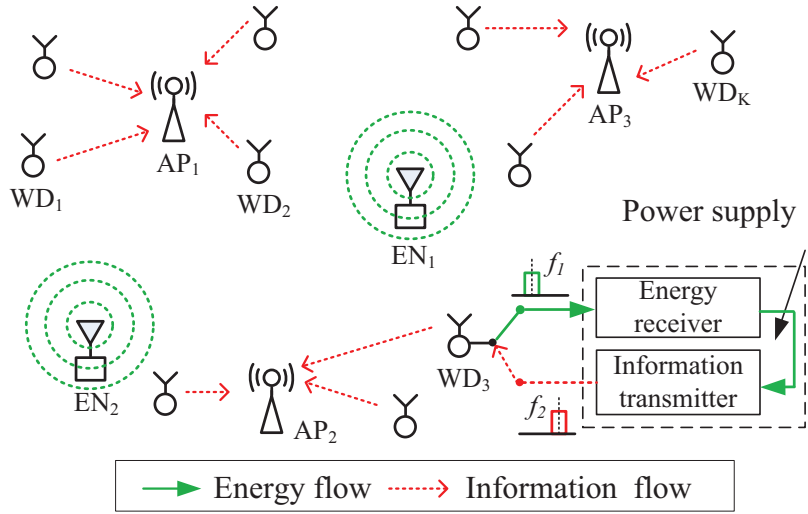}
  \end{center}
  \caption{Schematics of a WPCN with separate ENs and APs.}
  \label{61}
\end{figure}

For the case of separated ENs and APs, we consider in Fig.~\ref{61} a WPCN in $\mathbb{R}^2$ consisting of $M$ ENs, $N$ APs and $K$ WDs, whose locations are denoted by $2\times 1$ coordinate vectors $\left\{\mathbf{u}_i|i=1,\cdots,M\right\}$, $\left\{\mathbf{v}_j|j=1,\cdots,N\right\}$, and $\left\{\mathbf{w}_k |k=1,\cdots,K\right\}$, respectively. We assume that the energy and information transmissions are performed on orthogonal frequency bands without interfering with each other. Specifically, the ENs are connected to stable power source and broadcast RF energy in the DL for the WDs to harvest the energy and store in their rechargeable batteries. At the same time, the WDs use the battery power to transmit information to the APs in the UL. The circuit structure of a WD to perform the above operations is also shown in Fig.~\ref{61}.

In a transmission block of length $T$, the $M$ ENs transmit simultaneously on the same bandwidth in the DL, where each EN $i$ transmits
\begin{equation}
x_i(t) = \sqrt{P_0} s_{i}(t), \ t\in \left[0, T\right],\ i=1,\cdots,M.
\end{equation}
Here, $P_0$ denotes the transmit power, $s_{i}(t)$ denotes the pseudo-random energy signal used by the $i$-th EN, which is assumed to be of unit power ($\mathbb{E}_t\left[|s_{i}(t)|^2\right]=1$) and independent among the ENs ($\mathbb{E}_t\left[s_{i}(t)s_{j}(t)\right] = 0$ if $i\neq j$). The reason to use random signal instead of a single sinusoid tone is to avoid peak in transmit power spectrum density, for satisfying the equivalent isotropically radiated power (EIRP) requirement enforced by spectrum regulating authorities \cite{2014:Bi}. Notice that the energy beamforming technique proposed in \cite{2013:Zhang} is not used in our setup, as it requires accurate CSI and DL symbol-level synchronization, which may be costly to implement in a highly distributed WPCN network considered in this work.

Accordingly, the received energy signal by the $k$-th WD is
\begin{equation}
y_{k}(t) = \sqrt{P_0} \mathsmaller\sum_{i=1}^M \alpha_{i,k}  s_{i}(t), \ k=1,\cdots,K,
\end{equation}
where $\alpha_{i,k}$ denotes the equivalent baseband channel coefficient from the $i$-th EN to the $k$-th WD, which is assumed to be constant within a transmission block but may vary independently across different blocks. Let $h_{i,k}\triangleq |\alpha_{i,k}|^2$ denote the channel power gain, which follows a general distribution with its mean determined by the distance between the EN and WD, i.e.,
\begin{equation}
\label{3}
\mathbb{E}\left[h_{i,k}\right] =  \beta ||\mathbf{u}_i -\mathbf{w}_k||^{-d_D}, \ \ i=1,\cdots,M, \ k =1,\cdots K,
\end{equation}
where $d_D \geq 2$ denotes the path loss exponent in DL, $||\cdot||$ denotes the $l_2$-norm operator, and $\beta \triangleq A_d\left(\frac{3\cdot10^8}{4\pi f_d}\right)^{d_D}$ with $A_d$ and $f_d$ denoting the downlink antenna power gain and carrier frequency, respectively \cite{2005:Goldsmith}. Then, each WD $k$ can harvest an average amount of energy from the energy transmission within each block given by \cite{2013:Zhou}
\begin{equation}
\label{6}
\begin{aligned}
Q_{k} &= \eta T \mathbb{E} \left[|y_{k}(t)|^2|\left\{h_{i,k}\right\}\right] = \eta T P_0 \left( \mathsmaller\sum_{i=1}^M  h_{i,k}\right), \ k= 1,\cdots, K,
\end{aligned}
\end{equation}
where $\eta \in(0,1]$ denotes the energy harvesting circuit efficiency, and the expectation is taken over the pseudo-random energy signal variations under fixed $h_{i,k}$'s within the transmission block. Let $\lambda_k \triangleq \mathbb{E}\left[Q_k\right]/T$ denote the average energy harvesting rate over the variation of wireless channels ($h_{i,k}$'s) in different transmission blocks, we have
\begin{equation}
\label{10}
\lambda_k =  \eta \beta P_0 \cdot \mathsmaller\sum_{i=1}^M ||\mathbf{u}_i -\mathbf{w}_k||^{-d_D}, \ k=1,\cdots,K.
\end{equation}

In the UL information transmissions, we assume that each WD transmits data to only one of the APs. To make the placement problem tractable, the WD-AP associations are assumed to be fixed, where each WD $k$ transmits to its nearest AP $j_k$ regardless of the instantaneous CSI, i.e.,
\begin{equation}
\label{22}
j_k = \arg \min_{j=1,\cdots,N} ||\mathbf{v}_j - \mathbf{w}_k||,\ k=1\cdots,K.
\end{equation}
Here, we assume no co-channel interference for the received user signals from different WDs, e.g., the WDs transmit on orthogonal channels. Besides, for the simplicity of analysis, we assume no limit on the maximum number of WDs that an AP could receive data from. Then, the average power consumption rate for WD $k$ is modeled as
\begin{equation}
\label{11}
\mu_k = a_{1,k} + E_{k} \triangleq a_{1,k} + a_{2,k} ||\mathbf{v}_{j_k} - \mathbf{w}_k||^{d_U}, \ k=1,\cdots,K,
\end{equation}
where $a_{1,k}$ denotes the constant circuit power of WD $k$, $E_{k}$ denotes the average transmit power as a function of the distance between WD $k$ and its associated AP $j_k$, and $d_U\geq 2$ denotes the UL channel path loss exponent. Besides, $a_{2,k}>0$ denotes a parameter related to the transmission strategy used in the UL communication.\footnote{For example, given a receive signal power requirement $\Gamma_k$ to achieve a target data rate and maximum allowed outage portability $\psi_k$ for WD $k$, we have $a_{2,k}= \frac{\Gamma_k}{A_u}\left(\frac{4\pi f_u}{3\cdot 10^8}\right)^{d_U}\mathbf{E}_1\left(\ln\left(\frac{1}{1- \psi_k}\right)\right)$ when truncated channel inverse transmission \cite{2005:Goldsmith} is used under Rayleigh fading channel, where $A_u$ and $f_u$ denote the uplink antenna gain and carrier frequency, respectively, and $\mathbf{E}_1(x) \triangleq \int_{1}^{\infty} \frac{1}{t} \cdot e^{-t x }\  dt$ denotes the exponential integral function.} In general, the model in (\ref{11}) indicates that the transmit power increases as a polynomial function of the distance between the transmitter and receiver to satisfy certain communication quality requirement, e.g., minimum data rate or maximum allowed outage probability, which is widely used for wireless network performance analysis \cite{2005:Hou,2007:Liu}.

\begin{figure}
\centering
  \begin{center}
    \includegraphics[width=0.6\textwidth]{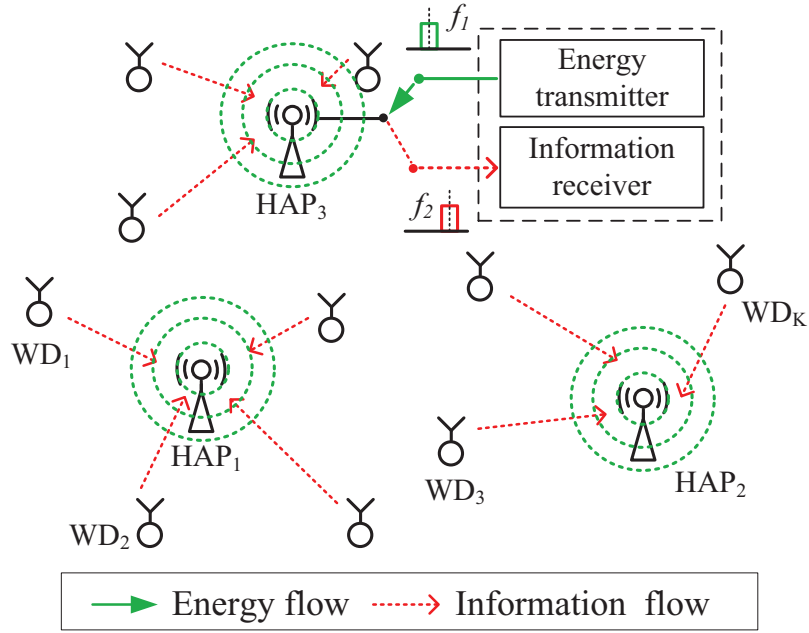}
  \end{center}
  \caption{Schematics of a WPCN with co-located ENs and APs (HAPs).}
  \label{62}
  \vspace{-5ex}
\end{figure}

\subsection{Co-located ENs and APs}
A special case of the WPCN that we consider in Fig.~\ref{61} is when the ENs and APs are grouped into pairs and each pair of EN and AP are co-located and integrated as a hybrid access point (HAP), which corresponds to setting $M=N$ and $\mathbf{u}_i =\mathbf{v}_i$ for $i=1,\cdots,M$. With the network model and HAP's circuit structure shown in Fig.~\ref{62}, a HAP transfers RF power in the DL and receives information in the UL simultaneously on different frequency bands. Although the use of HAPs is less flexible in placing the ENs and APs than with separated ENs and APs, the overall deployment cost is reduced, because the production and operation cost of a HAP is in general less than the sum-cost of two separate EN and AP.

For brevity, we reuse the notation $\mathbf{u}_i$, $i=1,\cdots,M$, to denote the location coordinates of the $M$ HAPs. Given other parameters unchanged, the expression of the average energy harvesting rate $\lambda_k$ of the $k$-th WD is the same as that in (\ref{10}). Meanwhile, the average power consumption rate $\mu_k$ can be obtained from (\ref{11}) by replacing $\mathbf{v}_j$ with $\mathbf{u}_i$ as follows.
\begin{equation}
\label{17}
\mu_k = a_{1,k} + a_{2,k} ||\mathbf{u}_{i_k} - \mathbf{w}_k||^{d_U}, \ k=1,\cdots,K,
\end{equation}
where $i_k$ is the index of the HAP that WD $k$ associates with, i.e.,
\begin{equation}
i_k = \arg \min_{i=1,\cdots,M} ||\mathbf{u}_i - \mathbf{w}_k||, \ k=1,\cdots,K.
\end{equation}

\subsection{Net Energy Harvesting Rate}
With the above definitions, the \textit{net} energy harvesting rates of the WDs in both cases of separate and co-located EN and AP are given by
\begin{equation}
\label{23}
\omega_k = \lambda_k - \mu_k,\ k=1,\cdots,K.
\end{equation}
In practice, the net energy harvesting rate can directly translate to the performance of device operating lifetime (see e.g., \cite{2015:Bi}). Specifically, given an initial battery level $C$, the average time before the $k$-th WD's battery depletes is $-C/\omega_k$ when $\omega_k<0$, and $+\infty$ when $\omega_k\geq 0$.\footnote{We neglect in this paper the battery degradation effect \cite{2013:Michelusi} caused by repeated charging and discharging.} In other words, given a minimum device operating lifetime requirement $T_k>0$, it must satisfy $\omega_k \geq -C/T_k$ if $T_k<\infty$, and $\omega_k\geq 0$ if $T_k = \infty$.

\section{Problem Formulation}\label{sec:PrbForm}
In this paper, we assume that the locations of the WDs are known and study the optimal placement of ENs and information APs, which are either separated or co-located in their locations. This may correspond to a sensor network with sensor (WD) locations predetermined by the sensed objects, or an IoT network with static WDs. In particular, we are interested in minimizing the deployment cost given that the net energy harvesting rates of all the WDs are larger than a common prescribed value $\gamma$, i.e., $\omega_k \geq \gamma,\ k=1\cdots,K$, where $\gamma$ is set to achieve a desired device operating lifetime.

\subsection{Separated ENs and APs}
When the ENs and APs are separated, the total deployment cost is $c_1 M+ c_2 N$ if $M$ ENs and $N$ APs are used, where $c_1$ and $c_2$ are the monetary costs of deploying an EN and an AP, respectively. To solve the minimum-cost deployment problem, let us first consider the following feasibility problem:
\begin{subequations}
\label{8}
   \begin{align}
    & \text{Find} & &  \mathbf{U}^M = \left[\mathbf{u}_1,\cdots,\mathbf{u}_M\right],\ \mathbf{V}^N = \left[\mathbf{v}_1,\cdots,\mathbf{v}_N\right]\\
    & \text{s. t.}    & & \lambda_k \left(\mathbf{U}^M\right) -\mu_k\left(\mathbf{V}^N\right) \geq \gamma, \ \ k=1,\cdots,K, \label{56}\\
    & & & \mathbf{b}^l\leq \mathbf{u}_i\leq \mathbf{b}^h,\ \ i=1,\cdots,M, \label{54}\\
    & & & \mathbf{b}^l\leq \mathbf{v}_j\leq \mathbf{b}^h,\ \ j=1,\cdots,N, \label{55}
   \end{align}
\end{subequations}
where $\lambda_k$ and $\mu_k$ are functions of $\mathbf{U}^M$ and $\mathbf{V}^N$ given in (\ref{10}) and (\ref{11}), respectively. The inequalities in (\ref{54}) and (\ref{55}) denote element-wise relations. Besides, $\left\{\mathbf{b}^l,\mathbf{b}^h\right\}$ specifies a feasible deployment area for both the ENs and APs in $\mathbb{R}^2$, which is large enough to contain all the WDs, i.e., $\mathbf{b}^l \leq \mathbf{w}_k \leq \mathbf{b}^h,\ k=1\cdots,K$. Evidently, if (\ref{8}) can be efficiently solved for any $M$ and $N$, then the optimal node placement solution to the considered minimum-cost deployment problem can be easily obtained through a simple two-dimension search over the values of $M$ and $N$, i.e., finding a pair of feasible $(M,N)$ that produces the lowest deployment cost $c_1 M+ c_2 N$.

For a pair of fixed $M$ and $N$ ($M>0$ and $N>0$), we can see that (\ref{8}) is feasible if and only if the optimal objective of the following problem is no smaller than $\gamma$, i.e.,
\begin{subequations}
\label{57}
   \begin{align}
    & \underset{\mathbf{U}^M,\mathbf{V}^N}{\text{max}} & \min_{k=1,\cdots,K}\ \  \left\{\lambda_k \left(\mathbf{U}^M\right) -\mu_k\left(\mathbf{V}^N\right)\right\}\\
    & \text{s. t.}  & \mathbf{b}^l\leq \mathbf{u}_i\leq \mathbf{b}^h,\ \ i=1,\cdots,M,\\
    & &  \mathbf{b}^l\leq \mathbf{v}_j\leq \mathbf{b}^h,\ \ j=1,\cdots,N.
   \end{align}
\end{subequations}
We can express (\ref{57}) as its equivalent epigraphic form \cite{2004:Boyd}, i.e.,
\begin{equation}
\label{9}
   \begin{aligned}
    &\underset{t,\mathbf{U}^M,\mathbf{V}^N}{\text{max}} & &  t\\
    &\text{s. t.}    & & \lambda_k \left(\mathbf{U}^M\right) -\mu_k\left(\mathbf{V}^N\right) \geq t, \ \ k=1,\cdots,K, \\
    & & & \mathbf{b}^l\leq \mathbf{u}_i\leq \mathbf{b}^h,\ \ i=1,\cdots,M,\\
    & & & \mathbf{b}^l\leq \mathbf{v}_j\leq \mathbf{b}^h,\ \ j=1,\cdots,N.
   \end{aligned}
\end{equation}
Given fixed $M$ and $N$, (\ref{8}) is feasible if and only if the optimal objective of (\ref{9}) satisfies $t^* \geq \gamma$. Then, the key difficulty of solving the optimal deployment problem is to find efficient solution for problem (\ref{9}).

\subsection{Co-located ENs and APs}
When the ENs and APs are integrated as HAPs, the total deployment cost is $c_3 M$ if $M$ HAPs are used. Here, $c_3$ denotes the cost of deploying a HAP, where in general $c_3<c_1+c_2$. Similar to the case of separated ENs and APs, the minimum-cost placement problem can be equivalently formulated as the following feasibility problem for any fixed number of $M>0$ HAPs,
\begin{equation}
\label{18}
   \begin{aligned}
    \text{Find}&  & &  \mathbf{U}^M = \left[\mathbf{u}_1,\cdots,\mathbf{u}_M\right]\\
    \text{s. t.}&     & &  \lambda_k \left(\mathbf{U}^M\right) -\mu_k\left(\mathbf{U}^M\right)\geq \gamma, \ \ k=1,\cdots,K,\\
    & & & \mathbf{b}^l\leq \mathbf{u}_i\leq \mathbf{b}^h,\ \ i=1,\cdots,M,
   \end{aligned}
\end{equation}
where $\lambda_k \left(\mathbf{U}^M\right)$ and $\mu_k\left(\mathbf{U}^M\right)$ are given in (\ref{10}) and (\ref{17}), respectively. Notice that the study on co-located ENs and APs is not a special case of that of separated ENs and APs. In fact, it adds extra constraints ($\mathbf{u}_i=\mathbf{v}_i,\ i=1,\cdots,M$) to (\ref{8}), which leave less flexibility to the nodes placement design and make the problem more challenging to solve. Equivalently, the feasibility of (\ref{18}) can be determined by solving the following optimization problem
\begin{equation}
\label{14}
   \begin{aligned}
    \underset{t,\mathbf{U}^M}{\text{max}}&  & & t \\
    \text{s. t.}&   &  & \lambda_k \left(\mathbf{U}^M\right) -\mu_k\left(\mathbf{U}^M\right) \geq t,\ \ k=1,\cdots,K, \\
    & & & \mathbf{b}^l\leq \mathbf{u}_i\leq \mathbf{b}^h,\ \ i=1,\cdots,M, \\
   \end{aligned}
\end{equation}
and then comparing the optimal objective $t^*$ with $\gamma$, to see whether $t^*\geq \gamma$ holds. In the following Sections IV and V, we propose efficient algorithms to solve problems (\ref{9}) and (\ref{14}), respectively. It is worth mentioning that the placement solution to (\ref{9}) and (\ref{14}) can be at arbitrary locations. When an EN (or a HAP) is placed at a location very close to an WD, the far-field channel model in (\ref{3}) may be inaccurate. However, we learn from (\ref{9}) and (\ref{14}) that the optimal value $t^*$ is determined by the performance-bottleneck WD that is far away from the ENs and APs (i.e., the channel model in (3) applies practically), thus having very low energy harvesting rate and high transmit power consumption. Therefore, the potential inaccuracy of (\ref{3}) will not affect the objective values of (\ref{9}) and (\ref{14}), and the proposed algorithms in this paper are valid in practice.

\section{Placement Optimization of Separated ENs and APs}
In this section, we study the node placement optimization for separately located EN and AP in problem (\ref{9}). Specifically, we first study in Section IV.A the method to optimize EN placement assuming that the locations of APs are fixed in a WPCN. In Section IV.B, we further study the method to optimize the placement of APs given fixed EN locations. Based on the obtained results, we then propose in Section IV.C an alternating method to jointly optimize the placements of ENs and APs. In addition, an alternative local searching method is considered in Section IV.D for performance comparison.

\subsection{EN Placement Optimization with Fixed AP Location}\label{sec:singEN}
We first consider the optimal EN placement problem when the locations of the APs are fixed, i.e., $\mathbf{v}_j$'s are known. In this case, the WD-AP association $j_k$ is known for each WD $k$ from (\ref{22}), and $\mu_k$'s can be calculated accordingly from (\ref{11}). It is worth mentioning that the proposed algorithms under the fixed AP setup can be directly extended to solve EN placement problem in other wireless powered networks not necessarily for communication purpose, e.g., a sensor network whose energy is mainly consumed on sensing and processing data, as long as the energy consumption rates $\mu_k$'s are known parameters. With $\mathbf{v}_j$'s and $\mu_k$'s being fixed, we can rewrite (\ref{9}) as
\begin{equation}
\label{13}
   \begin{aligned}
    \underset{t,\mathbf{U}^M}{\text{max}}&  & &  t\\
    \text{s. t.}&     & & \varphi\cdot \mathsmaller\sum_{i=1}^M ||\mathbf{u}_i -\mathbf{w}_k||^{-d_D} -\mu_k \geq t, \ \ k=1,\cdots,K,\\
    & & & \mathbf{b}^l\leq \mathbf{u}_i\leq \mathbf{b}^h,\ \ i=1,\cdots,M,
   \end{aligned}
\end{equation}
where $\varphi \triangleq  \eta \beta P_0$. We can see that (\ref{13}) is a non-convex optimization problem, because $||\mathbf{u}_i -\mathbf{w}_k||^{-d_D}$ is neither a convex nor concave function in $\mathbf{u}_i$. As it currently lacks of effective method to convert (\ref{13}) into a convex optimization problem, the optimal solution is in general hard to obtain. However, for a special case with $M=1$, i.e., placing only one EN, the optimal solution is obtained in the following. By setting $M=1$, (\ref{13}) can be rewritten as
\begin{subequations}
\label{15}
   \begin{align}
    \underset{t,\mathbf{u}_1}{\text{max}}&  & &  t\\
    \text{s. t.}&     & &  ||\mathbf{u}_1 - \mathbf{w}_k||^{d_D} \leq \frac{\varphi}{t+ \mu_k},  \ k=1,\cdots,K, \label{58}\\
    & & & \mathbf{b}^l\leq \mathbf{u}_1\leq \mathbf{b}^h.
   \end{align}
\end{subequations}
Although (\ref{15}) is still a non-convex optimization problem (as $\varphi/(t+ \mu_k)$ is not a concave function in $t$), it is indeed a convex feasibility problem over $\mathbf{u}_1$ when $t$ is fixed, which can be efficiently solved using the interior point method \cite{2004:Boyd}. Therefore, the optimal solution of (\ref{15}) can be obtained using a bi-section search method over $t$, whose pseudo-code is given in Algorithm \ref{46}. Notice that the right hand side (RHS) of (\ref{58}) is always positive during the bisection search over $t \in\left(LB_1,UB_1\right)$. Besides, we can infer that Algorithm \ref{46} converges to the optimal solution $t^*$, because problem (\ref{15}) is feasible for $t\leq  t^*$ and infeasible otherwise. The total number of feasibility tests performed is $\log_2\left[\left(UB_1-LB_1\right)/\sigma_1\right]$, where $\sigma_1$ is a predetermined parameter corresponding to a solution precision requirement.

\begin{algorithm}
\footnotesize
 \SetAlgoLined
 \SetKwData{Left}{left}\SetKwData{This}{this}\SetKwData{Up}{up}
 \SetKwRepeat{doWhile}{do}{while}
 \SetKwFunction{Union}{Union}\SetKwFunction{FindCompress}{FindCompress}
 \SetKwInOut{Input}{input}\SetKwInOut{Output}{output}
 \Input{WD locations $\mathbf{w}_k$'s, power consumption rates $\mu_k$'s}
 \Output{the optimal location of the EN $\mathbf{u}_1^*$}
 Initialize: $LB_1\leftarrow - \max_{k=1,\cdots, K}\mu_k$, \ \ $UB_1\leftarrow P_0$\;
 \Repeat{$|UB_1-LB_1|<\sigma_1$}{
      $t\leftarrow (UB_1+LB_1)/2$\;

      \eIf{Problem (\ref{15}) is feasible given $t$}{
      $LB_1 \leftarrow t$\;
      $\mathbf{\hat{u}}_1 \leftarrow$ a feasible solution of (\ref{15}) given $t$\;
      }{
      $UB_1 \leftarrow t$\;
      }

      }
\textbf{Return} $\mathbf{u}^* \leftarrow$ the last feasible solution $\mathbf{\hat{u}}_1$\;
\caption{Bi-section search for single EN placement.}
\label{46}
\end{algorithm}

Since placing one EN optimally is solved, we have the potential to decouple the difficult EN placement problem (\ref{13}) into $M$ relatively easy problems with $M>1$. This motivates a greedy algorithm, which places the ENs iteratively one-by-one into the network. Intuitively, an optimal deployment solution of (\ref{13}) should ``spread" the $M$ ENs among the $K$ WDs to maximize the minimum energy harvesting rate. However, the optimal solution obtained from solving (\ref{15}) tends to place the single EN around the center of the cluster formed by the $K$ WDs. Inspired by the optimal solution structures of (\ref{13}) and (\ref{15}), we propose in the following a cluster-based greedy EN placement method, where the newly placed EN optimizes the net energy harvesting rates of an expanding cluster of WDs, until all the WDs are included.

In practice, we geographically partition the $K$ WDs into $M$ non-overlapping clusters (assuming $K\geq M$), denoted by $\left\{\mathcal{W}_i| i=1,\cdots,M\right\}$.\footnote{The proposed node placement algorithms can apply to any clustering method used. Besides, the algorithm complexity is not related to the clustering method as long as the partitions of the WDs are given. For simplicity, we use $K$-means clustering algorithm in this paper.} This can be efficiently achieved by, e.g., the well-known $K$-means clustering algorithm \cite{2006:Bishop}. Then, in the $i$-th iteration ($i\geq 1$), we obtain the optimal location of the $i$-th EN, denoted by $\mathbf{u}_i^*$, by maximizing the net energy harvesting rates of the WDs in the first $i$ clusters as follows.
\begin{subequations}
\label{35}
   \begin{align}
    \underset{t_i,\mathbf{u}_i}{\text{max}}&  & &  t_i\\
    \text{s. t.}&     & & \left(t_i+ \mu_k - \lambda_{i-1,k}\right) \cdot ||\mathbf{u}_i - \mathbf{w}_k||^{d_D} \leq \varphi,  \nonumber \\
    & & & k\in \left\{\mathcal{W}_1 \cup \cdots \cup \mathcal{W}_i\right\}, \label{31}\\
    & & & \mathbf{b}^l\leq \mathbf{u}_i\leq \mathbf{b}^h,
   \end{align}
\end{subequations}
where $\lambda_{i-1,k}$ denotes the accumulative RF power harvested at the $k$-th WD due to the $(i-1)$ previously deployed ENs, given by
\begin{equation}
\label{53}
\lambda_{i-1,k} = \begin{cases}
0, & i=1,\\
\varphi \cdot \mathsmaller\sum_{j=1}^{i-1} ||\mathbf{u}_j^* - \mathbf{w}_k||^{-d_D}, & i>1.
\end{cases}
\end{equation}
In each iteration, $\mathbf{u}_i^*$ can be efficiently obtained using a bi-section search method over $t_i$ similar to that in solving (\ref{15}). The pseudo-code of the greedy algorithm is given in Algorithm \ref{41}. Notice in line $8$ of Algorithm \ref{41}, the corresponding inequality in (\ref{31}) holds automatically when $t_i + \mu_k - \bar{Q}_{i-1,k}<0$ for some $k$, given a fixed $t_i\in\left(LB_2,UB_2\right)$. Therefore, the corresponding constraints can be safely ignored without affecting the feasibility of (\ref{35}).

\begin{figure}
\centering
  \begin{center}
    \includegraphics[width=0.6\textwidth]{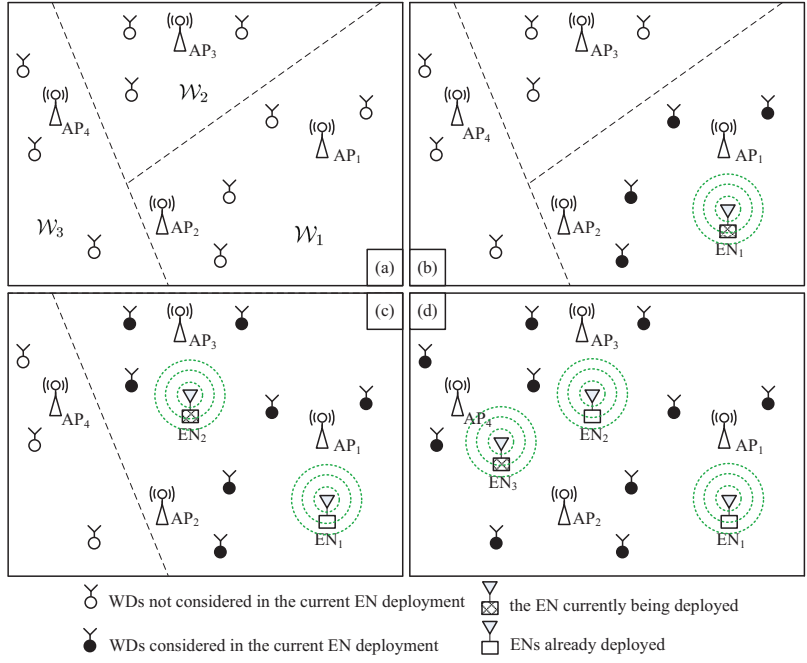}
  \end{center}
  \caption{Illustration of greedy algorithm for placing $M=3$ ENs (with fixed APs).}
  \label{63}
\end{figure}

The greedy EN placement method is illustrated in Fig. \ref{63}. In this example, we first divide the WDs into $M=3$ clusters in Fig. \ref{63}(a), then place the $3$ ENs one-by-one in Fig. \ref{63}(b)-(d). When placing the $1$st EN or EN$_1$, the algorithm only considers the received energy of the WDs in the $1$st cluster (shaded WDs in Fig. \ref{63}(b)) from the EN to be placed; for the $2$nd EN or EN$_2$, it considers the received energy of the WDs in the $1$st and $2$nd clusters from the first $2$ ENs; for the last EN or EN$_3$, it considers the received energy of all the WDs from the $3$ ENs. Notice that our greedy algorithm allows multiple ENs to be placed in the same cluster, because the placement of the $i$-th EN considers all the WDs in the first $i$ clusters.

Algorithm \ref{41} applies Algorithm \ref{46} $M$ times, one for placing each EN, thus the total number of feasibility tests performed is $M\log_2\left[\left(UB_2-LB_2\right)/\sigma_2\right]$, where $\sigma_2$ is a parameter corresponding to a solution precision requirement. Besides, the time complexity of solving each convex feasibility test using the interior point method is $O\left(\sqrt{K+2M}\log \left(K+2M\right)\right)$\cite{2004:Boyd}. Therefore, the overall time complexity of Algorithm \ref{41} is $O\left(M\sqrt{K+2M}\log \left(K+2M\right)\right)$, which is moderate even for a large-size network consisting of, e.g., tens of ENs and hundreds of WDs.

\begin{algorithm}
\footnotesize
\SetAlgoLined
 \SetKwData{Left}{left}\SetKwData{This}{this}\SetKwData{Up}{up}
 \SetKwRepeat{doWhile}{do}{while}
 \SetKwFunction{Union}{Union}\SetKwFunction{FindCompress}{FindCompress}
 \SetKwInOut{Input}{input}\SetKwInOut{Output}{output}
 \Input{WD locations $\mathbf{w}_k$'s, $N$ AP locations $\mathbf{v}_j$'s\;}
 \Output{locations of $M$ ENs $\left\{\mathbf{u}_1^*,\cdots,\mathbf{u}_M^*\right\}$\;}

 \textbf{Initialization}:
  Clustering the WDs into $\left\{\mathcal{W}_i, i=1,\cdots,M\right\}$ \;
  With $\mathbf{v}_j$'s, calculate $\mu_k$'s using (\ref{22}) and (\ref{11})\;
 \For{$i=1$ \KwTo $M$}{
   $LB_2\leftarrow - \max_{k=1,\cdots, K}\mu_k$, \ $UB_2\leftarrow MP_0$ \;
   Update $\left\{\bar{Q}_{i-1,k},\ k=1\cdots,K\right\}$ using (\ref{53})\;
   \Repeat{$|UB_2-LB_2|<\sigma_2$}{
      $t_i \leftarrow (UB_2+LB_2)/2$\;
      Ignore the constraints in (\ref{31}) with $t_i + \mu_k - \bar{Q}_{i-1,k}<0$\;

      \eIf{Problem (\ref{35}) is feasible given $t_i$}{
      $LB_2 \leftarrow t_i$\;
       $ \mathbf{u}_i^* \leftarrow$ a feasible solution of (\ref{35})\;
      }{
      $UB_2 \leftarrow t_i$\;
      }

      }
}
\textbf{Return} $\left\{\mathbf{u}_1^*,\cdots,\mathbf{u}_M^*\right\}$
\caption{Greedy algorithm for $M$ EN placement.}
\label{41}
\end{algorithm}

\subsection{AP Placement Optimization under Fixed ENs}
We then study in this subsection the method to optimize the placement of APs given fixed EN locations, i.e., $\mathbf{u}_i$'s are known. In this case, $\lambda_k$'s are fixed and can be calculated using (\ref{10}). With $\lambda_k$'s being fixed parameters, we can substitute (\ref{11}) into (\ref{9}) and formulate the optimal AP placement problem under fixed ENs as follows
\begin{equation}
\label{27}
   \begin{aligned}
    \underset{t,\mathbf{V}^N}{\text{max}}&    & & t \\
    \text{s. t.}&  & &\lambda_k - a_{1,k} - a_{2,k} ||\mathbf{v}_{j_k} - \mathbf{w}_k||^{d_U} \geq  t, \ k=1,\cdots,K,  \\
    &                   & &\mathbf{b}^l\leq \mathbf{v}_{j}\leq \mathbf{b}^h,\ \ j=1,\cdots,N,
   \end{aligned}
\end{equation}
where $j_k$ is the index of AP that WD $k$ associates with given in (\ref{22}). The above problem is non-convex because of the combinatorial nature of WD-AP associations, i.e., $j_k$'s are discrete indicators. However, notice that if $j_k$'s are known, (\ref{27}) is a convex problem that is easily solvable. In practice, however, $j_k$'s are revealed only after (\ref{27}) is solved and the placement of APs is obtained. To resolve this conflict, we propose in the following a \emph{trial-and-error} method to find feasible $j_k$'s and accordingly a feasible AP placement solution to (\ref{27}). The pseudo-code of the method to solve (\ref{27}) is presented in Algorithm \ref{42} and explained as follows.

As its name suggests, we first convert (\ref{27}) into a convex problem by assuming a set of WD-AP associations, denoted by $j_k^{(b)}$, $k=1,\cdots,K$, and then solve (\ref{27}) for the optimal AP placement based on the assumed $j_k^{(b)}$'s. Next, we compare $j_k^{(b)}$'s with the actual WD-AP associations after the optimal AP placement is obtained using (\ref{22}), denoted by  $j_k^{(a)}$, $k=1,\cdots,K$. Specifically, we check if $j_k^{(a)} = j_k^{(b)},\forall k$. If yes, we have obtained a feasible solution to (\ref{27}); otherwise, we update $j_k^{(b)} = j_k^{(a)}$, $k=1,\cdots,K$ and repeat the above process until $j_k^{(a)} = j_k^{(b)}$, $\forall k$. The convergence of Algorithm \ref{42} is proved in the Appendix and the convergence rate is evaluated numerically in Fig.~\ref{64} of Section VI. Intuitively, the trial-and-error method is convergent because the optimal value of (\ref{27}) is bounded, while by updating $j_k^{(b)} = j_k^{(a)}$, we can always improve the optimal objective value of (\ref{27}) in the next round of solving it. As we will show later in Fig.~\ref{64} of Section VI, the number of iterations used until convergence is of constant order, i.e., $O(1)$, regardless of the value of $N$ or $K$. There, the time complexity of Algorithm \ref{42} is $O\left(\sqrt{K+2N}\log \left(K+2N\right)\right)$, as it takes this time complexity for solving (\ref{27}) in each iteration.

\begin{algorithm}
\footnotesize
 \SetAlgoLined
 \SetKwData{Left}{left}\SetKwData{This}{this}\SetKwData{Up}{up}
 \SetKwRepeat{doWhile}{do}{while}
 \SetKwFunction{Union}{Union}\SetKwFunction{FindCompress}{FindCompress}
 \SetKwInOut{Input}{input}\SetKwInOut{Output}{output}
 \Input{$K$ WD locations $\mathbf{w}_k$'s, $M$ EN locations $\mathbf{u}_i$'s\; }
 \Output{locations of $N$ APs, i.e., $\left\{\mathbf{v}_1^*,\cdots,\mathbf{v}_N^*\right\}$\;}

 \textbf{Initialization}:
 \\ Separate the WDs into $N$ clusters, and place each AP at a cluster center. Use $\mathbf{v}_j^{(0)}$'s to denote the initial AP locations\;
 With $\mathbf{v}_j^{(0)}$'s, calculate $j_k^{(b)}$'s using (\ref{22})\;
 With $\mathbf{u}_i$'s, calculate $\lambda_k$'s using (\ref{10}). Let $flag \leftarrow 1$\;

  \While{$flag =1$}{
  Given $j_k^{(b)}$'s, solve (\ref{27}) for optimal AP placement $\mathbf{v}_j^*$'s\;
  Given $\mathbf{v}_j^*$'s, calculate $j_k^{(a)}$'s using (\ref{22})\;

    \eIf{$j_k^{(a)}\neq j_k^{(b)}$ for some $k$}{
    Update $j_k^{(b)}= j_k^{(a)}$, $k=1,\cdots,K$\;
  }{
    A local optimum is found, return $\left\{\mathbf{v}_1^*,\cdots,\mathbf{v}_N^*\right\}$\;
    $flag \leftarrow 0$\;
  }

  }
\caption{Trial-and-error method for $N$ AP placement}
\label{42}
\end{algorithm}

\subsection{Joint EN and AP Placement Optimization}
In this subsection, we further study the problem of joint EN and AP placement optimization. In this case, we consider both the locations of ENs and APs as variables, such that the joint EN-AP placement problem in (\ref{9}) can be expressed as
\begin{equation}
\label{16}
   \begin{aligned}
    & \underset{t,\mathbf{U}^M,\mathbf{V}^N}{\text{max}} & &  t\\
    & \text{s. t.}  & & \varphi\cdot \mathsmaller\sum_{i=1}^M ||\mathbf{u}_i -\mathbf{w}_k||^{-d_D} - a_{2,k} ||\mathbf{v}_{j_k} - \mathbf{w}_k||^{d_U}\\
    & & & \geq t + a_{1,k}, \ \ k=1,\cdots,K,\\
    & & & \mathbf{b}^l\leq \mathbf{u}_i\leq \mathbf{b}^h,\ i=1,\cdots,M \\
    & & & \mathbf{b}^l\leq \mathbf{v}_j\leq \mathbf{b}^h,\ j=1,\cdots,N.
   \end{aligned}
\end{equation}
Evidently, the optimization problem is highly non-convex because of the non-convex function $||\mathbf{u}_i -\mathbf{w}_k||^{-d_D}$ and the discrete variables $j_k$'s. Based on the results in Section IV.A and IV.B, we propose an alternating method in Algorithm \ref{43} to solve (\ref{16}) for joint EN and AP placement solution. Specifically, starting with a feasible AP placement, we alternately apply Algorithms \ref{41} and \ref{42} to iteratively update the locations of ENs and APs, respectively. A point to notice is that Algorithm \ref{41} (and Algorithm \ref{42}) only produces a sub-optimal solution to (\ref{13}) (and (\ref{27})), thus the objective value of (\ref{16}) may decrease during the alternating iterations. To cope with this problem, we record the deployment solutions obtained in $L>1$ iterations and select the one with the best performance. The impact of the parameter $L$ to the algorithm performance is evaluated in Fig.~\ref{68} of Section VI. Given the complexities of Algorithms \ref{41} and \ref{42}, we can easily infer that the time complexity of Algorithm \ref{43} is $O\left(LM\sqrt{K+2M}\log \left(K+2M\right)+ L\sqrt{K+2N}\log \left(K+2N\right) \right)$.

\begin{algorithm}
\footnotesize
 \SetAlgoLined
 \SetKwData{Left}{left}\SetKwData{This}{this}\SetKwData{Up}{up}
 \SetKwRepeat{doWhile}{do}{while}
 \SetKwFunction{Union}{Union}\SetKwFunction{FindCompress}{FindCompress}
 \SetKwInOut{Input}{input}\SetKwInOut{Output}{output}
 \Input{$K$ WD locations $\mathbf{w}_k$'s, $L$ iterations\; }
 \Output{Locations of $M$ ENs and $N$ APs, i.e., $\mathbf{u}_i$'s and $\mathbf{v}_j$'s.\;}

 \textbf{Initialize}: Separate the WDs into $N$ clusters, and place each AP at a cluster center. Use $\mathbf{v}_j$'s to denote the initial AP locations\;

  \For{$l=1$ \KwTo $L$}{

  \eIf{$l$ is odd}{
  Given $\mathbf{v}_j$'s, solve (\ref{16}) for $\mathbf{u}_i$'s using Algorithm 2\;
  }{
  Given $\mathbf{u}_i$'s, solve (\ref{16}) for $\mathbf{v}_j$'s using Algorithm 3\;
  }
  $z_l \leftarrow \min_{k=1,\cdots,K}\left(\lambda_k - \mu_k\right)$, where $\lambda_k$ and $\mu_k$ are in (\ref{10}) and (\ref{11}), respectively.\;
  $\mathbf{u}_i^{(l)}\leftarrow \mathbf{u}_i,\ i=1,\cdots, M$, and $\mathbf{v}_j^{(l)}\leftarrow \mathbf{v}_j,\ j=1,\cdots, N$.
  }
$p \leftarrow \arg \max_{l=1,\cdots,L} z_l$\;
\textbf{Return:} $\mathbf{u}_i^{(p)}$'s and $\mathbf{v}_j^{(p)}$'s.

\caption{An alternating method for joint AP-EN placement.}
\label{43}
\end{algorithm}

\subsection{Alternative Method}
Besides the proposed alternating method for solving (\ref{16}), we also consider an alternative local searching method used as benchmark algorithm for performance comparison. The local searching algorithm starts with a random deployment of the $M$ ENs and $N$ APs, i.e., $\mathbf{u}_i$'s and $\mathbf{v}_i$'s, and checks if the minimum net energy harvesting rate among the WDs, i.e.,
\begin{equation}
\begin{aligned}
P_r \triangleq \min_{k = 1,\cdots,K} &\biggl(\varphi \cdot \mathsmaller\sum_{i=1}^M ||\mathbf{u}_i -\mathbf{w}_k||^{-d_D}  - a_{1,k}- a_{2,k} \min_{j=1,\cdots,N} ||\mathbf{v}_{j} - \mathbf{w}_k||^{d_U}\biggr)
\end{aligned}
\end{equation}
can be increased by making a random movement to $\mathbf{\bar{u}}_i$'s and $\mathbf{\bar{v}}_j$'s that satisfy
\begin{equation}
\begin{aligned}
&\biggl\{\mathbf{\bar{u}}_i,\mathbf{\bar{v}}_j, i=1,\cdots,M,\ j=1,\cdots,N, \bigg| \mathsmaller\sum_{i=1}^M ||\mathbf{\bar{u}}_i-\mathbf{u}_i||^2 + \mathsmaller\sum_{j=1}^N ||\mathbf{\bar{v}}_j-\mathbf{v}_j||^2< \sigma_3\biggr\},
\end{aligned}
\end{equation}
where $\sigma_3$ is a fixed positive parameter. If yes, it makes the move and repeats the random movement process. Otherwise, if $P_r$ cannot be increased, the algorithm has reached a local maximum and returns the current placement solution. Several off-the-shelf local searching algorithms are available, where simulated annealing \cite{2010:Hromkovic} is used in this paper. In particular, simulated annealing can improve the searching result by allowing the nodes to be moved to locations with decreased value of $P_r$ to reduce the chance of being trapped at local maximums. Besides, we can improve the quality of deployment solution using different initial node placements, which are obtained either randomly or empirically, and select the resulted solution with the best performance.

\section{Placement Optimization of Co-located ENs and APs}
In this section, we proceed to study the node placement optimization problem (\ref{14}) for the case of co-located ENs and APs. The problem is still non-convex due to which the optimal solution is hard to be obtained. Inspired by both Algorithms \ref{41} and \ref{42}, we propose in this section an efficient greedy algorithm for HAP placement optimization.

\subsection{Greedy Algorithm Design}
The node placement optimization problem (\ref{14}) is highly non-convex, because the expression of problem (\ref{14}) involves non-convex function $||\mathbf{u}_i -\mathbf{w}_k||^{-d_D}$ in $\lambda_k\left(\mathbf{U}^M\right)$ and minimum operator over convex functions in $\mu_k\left(\mathbf{U}^M\right)$. Since its optimal solution is hard to obtain, a promising alternative is the greedy algorithm, which iteratively places a single HAP to the network at one time, similar to Algorithm \ref{41} for solving (\ref{13}) which optimizes the EN locations given fixed APs. However, by comparing problems (\ref{14}) and (\ref{13}), we can see that the algorithm design for solving (\ref{14}) is more complicated, because each $\mu_k$ is now a function of $\mathbf{u}_i$'s, instead of constant parameter in (\ref{13}).

Similar to the greedy algorithm in Section IV.A, we first separate the $K$ WDs into $M$ non-overlapping clusters, denoted by $\left\{\mathcal{W}_i| i=1,\cdots,M\right\}$, and add to the network a HAP in each iteration. Specifically, in the $i$-th iteration, given that the previous $(i-1)$ HAPs are fixed, we obtain the optimal location of the $i$-th HAP, denoted by $\mathbf{u}_i^*$, by maximizing the net energy harvesting rates of the WDs in the first $i$ clusters. To simplify the notations, we also use $\lambda_{i-1,k}$ as in Section IV.A to denote the accumulative RF harvesting power of the WD $k$ from the previously placed $(i-1)$ HAPs, which can be calculated using (\ref{53}). Besides, let $\mu_{i-1,k}$ denote the energy consumption rate of the $k$-th WD after the first $(i-1)$ HAPs have been placed, where
\begin{equation}
\label{52}
\mu_{i-1,k} = \begin{cases}
+\infty, & i=1,\\
a_{1,k} + a_{2,k} \min_{j=1,\cdots,i-1} ||\mathbf{u}_{j}^* - \mathbf{w}_k||^{d_U}, & i>1.
\end{cases}
\end{equation}
Notice that the only difference between placing the $i$-th HAP and the $i$-th EN in Section IV.A is that $\mu_k$ is now a function of $\mathbf{u}_i$ instead of a given constant. By substituting (\ref{52}) into (\ref{35}), the optimal location of the $i$-th HAP is obtained by solving the following problem
\begin{subequations}
\label{49}
   \begin{align}
    & \underset{t_i,\mathbf{u}_i}{\text{max}} & &  t_i\\
    & \text{s. t.}    & & \left(t_i + \mu_{i,k} - \lambda_{i-1,k}\right) \cdot ||\mathbf{u}_i - \mathbf{w}_k||^{d_D} \leq \varphi, \nonumber \\
    &  & & k \in \left\{\mathcal{W}_1\cup \cdots \cup \mathcal{W}_i\right\}, \label{51}\\
    & & & \mathbf{u}^l\leq \mathbf{u}_i\leq \mathbf{u}^h,
   \end{align}
\end{subequations}
where
\begin{equation}
\label{32}
\mu_{i,k} = \min\left(\mu_{i-1,k}, a_{1,k} + a_{2,k} ||\mathbf{u}_{i} - \mathbf{w}_k||^{d_U}\right).
\end{equation}
From (\ref{32}), we can see that a WD may change its association to the $i$-th HAP, if the newly placed HAP is closer to the WD than all the other $(i-1)$ HAPs that have been previously deployed. This combinatorial nature of WD-AP associations makes problem (\ref{49}) non-convex even if $t_i$ is fixed. In the following, we apply the similar trial-and-error technique as that in Section IV.B to obtain a feasible solution to problem (\ref{49}).

\subsection{Solution to Problem (\ref{49})}
The basic idea to obtain a feasible solution of (\ref{49}) is to convert it into a convex problem given $t_i$, and then use simple bi-section search over $t_i$. The convexification of (\ref{49}) is achieved by a trial-and-error method similar to that used for finding feasible WD-AP associations proposed in Algorithm \ref{42}. That is, we iteratively make assumptions on WD-AP associations and update the optimal placement of the $i$-th HAP obtained from solving (\ref{49}) based on the assumptions in the current iteration. With a bit abuse of notations, here we reuse $\mathbf{u}_i^*$  in each iteration as the optimal location of the $i$-th HAP given the current WD-AP association assumptions. Specifically, we assume whether the WDs change their associations after the $i$-th HAP is added, i.e., assuming either $\mu_{i-1,k}< a_{1,k} + a_{2,k} ||\mathbf{u}_{i}^* - \mathbf{w}_k||^{d_U}$ or $\mu_{i-1,k} \geq a_{1,k} + a_{2,k} ||\mathbf{u}_{i}^* - \mathbf{w}_k||^{d_U}$ for each $k$. Then, given a fixed $t_i$, each constraint on $k$ in (\ref{51}) belongs to one of the following four cases:

\subsubsection{Case $1$}
If we assume that WD $k$ does not change its WD-HAP association after the $i$-th HAP is placed into the WPCN, or equivalently $\mu_{i-1,k}< a_{1,k} + a_{2,k} ||\mathbf{u}_{i}^* - \mathbf{w}_k||^{d_U}$, we can replace the corresponding constraint in (\ref{51}) with
\begin{equation}
\label{47}
\left(t_i+ \mu_{i-1,k} - \lambda_{i-1,k}\right)\cdot||\mathbf{u}_i - \mathbf{w}_k||^{d_D} \leq \varphi.
\end{equation}
With a fixed $t_i$, (\ref{47}) is a convex constraint if $t_i+ \mu_{i-1,k} - \lambda_{i-1,k} > 0$.

\subsubsection{Case $2$} If we still assume $\mu_{i-1,k}< a_{1,k} + a_{2,k} ||\mathbf{u}_{i}^* - \mathbf{w}_k||^{d_U}$, while $t_i+ \mu_{i-1,k} - \lambda_{i-1,k} \leq 0$ holds, we can safely drop the constraint in (\ref{51}) without changing the feasible region of $\mathbf{u}_i$.

\subsubsection{Case $3$} On the other occasion, if we assume that WD $k$ changes its WD-HAP association, or $\mu_{i-1,k}\geq a_{1,k} + a_{2,k} ||\mathbf{u}_{i}^* - \mathbf{w}_k||^{d_U}$, the corresponding constraint in (\ref{51}) becomes
\begin{equation}
\label{45}
\left(t_i + a_{1,k} + a_{2,k}||\mathbf{u}_i-\mathbf{w}_k||^{d_U} - \lambda_{i-1,k}\right)\cdot ||\mathbf{u}_i - \mathbf{w}_k||^{d_D} \leq \varphi,
\end{equation}
which can be further expressed as
\begin{equation}
\label{33}
||\mathbf{u}_i-\mathbf{w}_k||^{d_U+ d_D} + \frac{t_i+ a_{1,k}- \lambda_{i-1,k}}{a_{2,k}}||\mathbf{u}_i - \mathbf{w}_k||^{d_D} -\frac{\varphi}{a_{2,k}} \leq 0.
\end{equation}
Notice that, given a fixed $t_i$, (\ref{33}) is a convex constraint if $t_i+ a_{1,k}- \lambda_{i-1,k} \geq 0$.

\subsubsection{Case $4$}
Otherwise, if we assume $\mu_{i-1,k}\geq a_{1,k} + a_{2,k} ||\mathbf{u}_{i}^* - \mathbf{w}_k||^{d_U}$ and $t_i+ a_{1,k}- \lambda_{i-1,k} < 0$ holds, (\ref{33}) is a non-convex constraint, as the left-hand-side (LHS) of (\ref{33}) is the difference of two convex functions. Nonetheless, we show that (\ref{33}) can still be converted into a convex constraint in this case. Let us first consider a function
\begin{equation}
\label{36}
z(x) = x^{d_U+ d_D} + \frac{t_i + a_{1,k}- \lambda_{i-1,k}}{a_{2,k}}x^{d_D} -\frac{\varphi}{a_{2,k}},
\end{equation}
where $x\geq 0$ and $t_i + a_{1,k}- \lambda_{i-1,k}<0$. We calculate the first order derivative of $z(x)$ and find that $z$ increases monotonically when
\begin{equation}
\label{34}
x > \left[\frac{-\left(t_i + a_{1,k}- \lambda_{i-1,k}\right)d_D}{a_{2,k}\left(d_U+d_D\right)}\right]^{1/d_U} \triangleq \tau_{i,k},
\end{equation}
and decreases monotonically if $x\leq \tau_{i,k}$. Notice that $\tau_{i,k}>0$ and $z(0) = -\frac{\varphi}{a_{2,k}} <0$ always hold. Therefore, although $z(x)$ is not a convex function, $z(x)<0$ can still be equivalently expressed as $x<\theta_{i,k}$, with $\theta_{i,k}$ being some positive number satisfying $z(\theta_{i,k})=0$. The value of $\theta_{i,k}$ can be efficiently obtained using many off-the-shelf numerical methods, such as the classic Newton's method or bi-section search method. A close comparison between the LHS of (\ref{33}) and $z(x)$ in (\ref{36}) shows that, by letting $x \triangleq ||\mathbf{u}_i-\mathbf{w}_k||$, we can equivalently express (\ref{33}) as a convex constraint
\begin{equation}
\label{37}
||\mathbf{u}_i-\mathbf{w}_k|| \leq \theta_{i,k},
\end{equation}
when $t_i + a_{1,k}- \lambda_{i-1,k}<0$ holds.

To sum up, given a fixed $t_i$, we tackle the $k$-th constraint in (\ref{51}) using one of the following methods:
\begin{enumerate}
  \item Replace by (\ref{47}) if assuming $\mu_{i-1,k}< a_{1,k} + a_{2,k} ||\mathbf{u}_{i}^* - \mathbf{w}_k||^{d_U}$ and $t_i+ \mu_{i-1,k} - \lambda_{i-1,k} > 0$;
  \item Drop the constraint if assuming $\mu_{i-1,k}< a_{1,k} + a_{2,k} ||\mathbf{u}_{i}^* - \mathbf{w}_k||^{d_U}$ and $t_i+ \mu_{i-1,k} - \lambda_{i-1,k} \leq 0$;
  \item Replace by (\ref{33}) if assuming $\mu_{i-1,k}\geq a_{1,k} + a_{2,k} ||\mathbf{u}_{i}^* - \mathbf{w}_k||^{d_U}$ and $t_i+ a_{1,k}- \lambda_{i-1,k} \geq 0$;
  \item Replace by (\ref{37}) if assuming $\mu_{i-1,k}\geq a_{1,k} + a_{2,k} ||\mathbf{u}_{i}^* - \mathbf{w}_k||^{d_U}$ and $t_i + a_{1,k}- \lambda_{i-1,k}< 0$.
\end{enumerate}
After processing all the $K$ constraints in (\ref{51}), we can convert (\ref{49}) into a convex feasibility problem given a set of WD-HAP association assumptions and a fixed $t_i$. Accordingly, the optimal placement of the $i$-th HAP ($\mathbf{u}_i^*$) under the assumptions, can be efficiently obtained from solving (\ref{49}) using a bi-section search method over $t_i$. Similar to the trial-and-error technique used in Algorithm \ref{42}, we check if the obtained $\mathbf{u}_i^*$ satisfies all the assumptions made. If yes, we have obtained a feasible solution of (\ref{49}). Otherwise, we switch the violating assumptions, then follow the above constraint processing method to resolve (\ref{49}) for a new $\mathbf{u}_i^*$, and repeat the iterations until all the assumptions are satisfied. The above trial-and-error method converges. The proof follows the similar argument as given in the Appendix, which proves the convergence of the trial-and-error method used for solving problem (\ref{27}). Thus, this proof is omitted here.

\begin{figure}
\centering
  \begin{center}
    \includegraphics[width=0.6\textwidth]{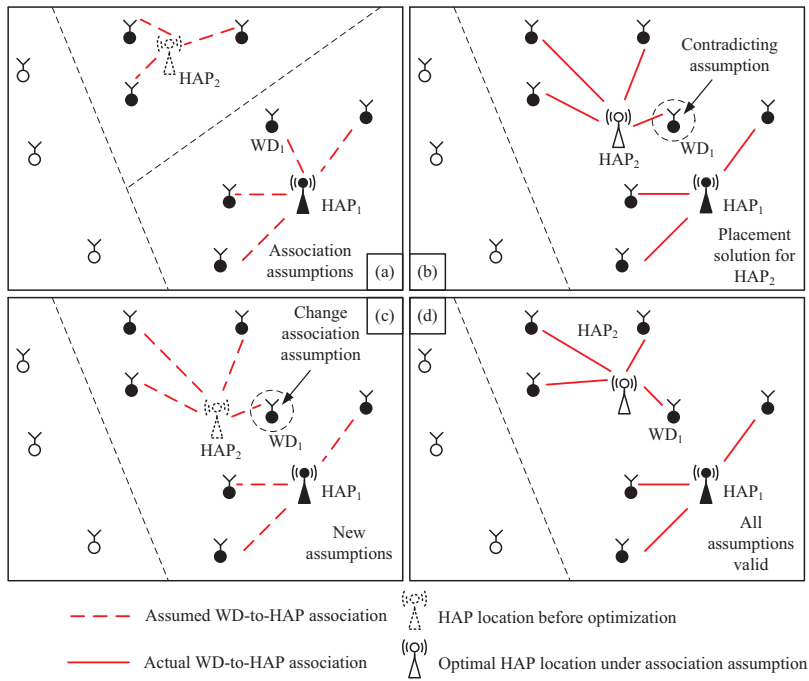}
  \end{center}
  \caption{Illustration of greedy algorithm for placement optimization with co-located EN and AP (HAP).}
  \label{66}
\end{figure}

\subsection{Overall Algorithm}
Since the placement of a single HAP can be obtained via solving (\ref{49}), we can iteratively place the $M$ HAPs into the WPCN. The pseudo-code of the revised greedy algorithm is presented in Algorithm \ref{44}. For example, Fig. \ref{66} illustrates the detailed steps taken to place the $2$nd HAP, or HAP$_2$ (total $3$ HAPs while the first HAP, or HAP$_1$ is already placed). Specifically, we first assume in Fig. \ref{66}(a) that all the WDs in the $1$st cluster associate with HAP$_1$, and the WDs in the $2$nd cluster associate with HAP$_2$, after HAP$_2$ is added into the network. Then, we obtain in Fig. \ref{66}(b) the optimal placement of the HAP$_2$ based on the association assumptions made. However, the obtained location of HAP$_2$ results in a contradiction with the association assumption made on WD$_1$ (assumed to be associated with HAP$_1$). Therefore, we change the association assumption of WD$_1$ to HAP$_2$, and recalculate the optimal placement solution for HAP$_2$ (Fig. \ref{66}(c)). In Fig. \ref{66}(d), the newly obtained location of HAP$_2$ satisfies all the association assumptions, thus the placement of HAP$_2$ is feasible. Following the similar argument in the Appendix, the association assumption update procedure converges, because the optimal objective value of (\ref{49}) is non-decreasing upon each association assumptions update (lines $10-27$ of Algorithm \ref{44}). After obtaining the location of HAP$_2$, a feasible location of the $3$rd HAP, HAP$_3$, can also be obtained using the similar procedures as above. Besides, we can infer that the time complexity of Algorithm \ref{44} is $O\left(M\sqrt{K+2M}\log\left(K+2M\right)\right)$, because it places the $M$ HAPs iteratively, while the trial-and-error method used to place each HAP needs $O\left(\sqrt{K+2M}\log\left(K+2M\right)\right)$ complexity.

\begin{algorithm}
\footnotesize
 \SetAlgoLined
 \SetKwData{Left}{left}\SetKwData{This}{this}\SetKwData{Up}{up}
 \SetKwRepeat{doWhile}{do}{while}
 \SetKwFunction{Union}{Union}\SetKwFunction{FindCompress}{FindCompress}
 \SetKwInOut{Input}{input}\SetKwInOut{Output}{output}
 \Input{$K$ WD locations $\mathbf{w}_k$'s}
 \Output{locations of $M$ HAPs $\left\{\mathbf{u}_1^*,\cdots,\mathbf{u}_M^*\right\}$}
  Cluster the WDs into $\left\{\mathcal{W}_i, i=1,\cdots,M\right\}$\;
 \For{$i=1$ \KwTo $M$}{
 \For{ each WD $k$}{
       Update $\lambda_{i-1,k}$ and $\mu_{i-1,k}$ using (\ref{53}) and (\ref{52})\;
       Assume WD $k$ satisfies condition $(a)$ or $(b)$:\\
        \ \ $(a)$  $\mu_{i-1,k}< a_{1,k} + a_{2,k} ||\mathbf{u}_{i}^* - \mathbf{w}_k||^{d_U}$\;
        \ \ $(b)$ $\mu_{i-1,k} \geq a_{1,k} + a_{2,k} ||\mathbf{u}_{i}^* - \mathbf{w}_k||^{d_U}$.
 }
  $StopFlag \leftarrow 0$\;
   \Repeat{$StopFlag = 1$}{
   $LB\leftarrow -\delta$, \ $UB\leftarrow \delta$,\ \ $\delta$ is sufficiently large\;
    \Repeat{$|UB-LB|<\sigma$, $\sigma$ is sufficiently small}{
    $t_i \leftarrow (UB+LB)/2$\;
      Given $t_i$, convert (\ref{49}) into a convex problem using the procedures in Section V.B\;
    \eIf{Problem (\ref{49}) is feasible given $t_i$}{
      $LB \leftarrow t_i$; $ \mathbf{u}_i^* \leftarrow$ a feasible solution of (\ref{49})\;
      }{
      $UB \leftarrow t_i$\;
      }
    }
         \eIf{all the $K$ assumptions are valid}{
      $StopFlag \leftarrow 1$; the $i$-th HAP location $\leftarrow \mathbf{u}_i^*$\;
      }{
      $StopFlag \leftarrow 0$\;
      For each WD violating the assumption, switch the assumption from $(a)$ to $(b)$, or $(b)$ to $(a)$\;
      }

      }

 }
\textbf{Return} the HAP locations $\left\{\mathbf{u}_1^*,\cdots,\mathbf{u}_M^*\right\}$.
\caption{Greedy algorithm for HAP placement optimization.}
\label{44}
\end{algorithm}

\section{Simulation Results}
In this section, we use simulations to evaluate the performance of the proposed node placement methods. All the computations are executed by MATLAB on a computer with an Intel Core i5 $2.90$-GHz CPU and $4$ GB of memory. The carrier frequency is $915$ MHz for both DL and UL transmissions operating on different bandwidths. In the DL energy transmission, we consider using Powercast TX91501-1W power transmitter with $P_0=1$W (Watt) transmit power, and P2110 receiver with $A_d = 3$ dB antenna gain and $\eta= 0.51$ energy harvesting efficiency. Besides, we assume the path loss exponent $d_D = 2.2$, thus $\beta= 6.57\times 10^{-4}$. In the UL information transmission, we assume that $d_U = 2.5$, $a_{1,k} = 50\mu$W and $a_{2,k} = 1.4\cdot 10^{-6}$ for $k=1,\cdots,K$, where $a_{2,k}$ is obtained assuming Rayleigh fading and the use of truncated channel inversion transmission \cite{2005:Goldsmith} with receiver signal power $-70$dBm (equivalent to $18$dB SNR target with -$88$dBm noise power) and outage probability of $5\%$. All the WDs, ENs and APs are placed within a $24 m\times 24 m$ box region specified by $\mathbf{b}^l=(0,0)^T$ and $\mathbf{b}^h = (24,24)^T$. Unless otherwise stated, each point in the following figures is an average performance of $20$ random WD placements, each with $K= 60$ WDs uniformly placed within the box region.

\subsection{Separated EN and AP Deployment}
\begin{figure}
\centering
  \begin{center}
    \includegraphics[width=0.6\textwidth]{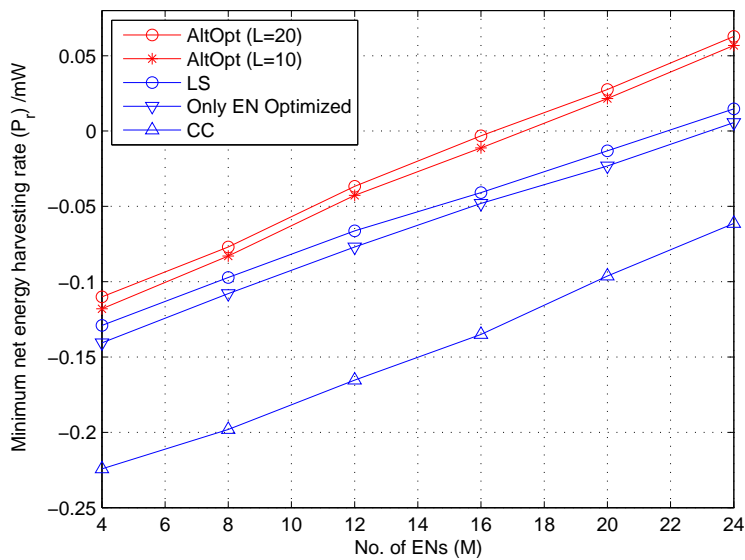}
  \end{center}
  \caption{Performance comparison of the separated AP and EN placement methods ($N=6$).}
  \label{68}
\end{figure}

We first evaluate the performance of the proposed alternating optimization method (Algorithm \ref{43}) for placing separated ENs and APs. Without loss of generality, we consider $N=6$ APs and show in Fig. \ref{68} the minimum net energy harvesting rate $P_r$ in (\ref{14}) achieved by Algorithm \ref{43} when the locations of APs are jointly optimized with those of different number of ENs ($M$). Evidently, a larger $P_r$ indicates better system performance. For the proposed alternating optimization algorithm (AltOpt), we show both the performance with $L=10$ and $20$. Besides, we also consider the following benchmark placement methods
\begin{itemize}
  \item Cluster center method (CC): separate the WDs into $M$ clusters and place an EN at each of the cluster centers. Similarly, separate the WDs into $N$ clusters and place the $N$ APs at the cluster centers;
  \item Optimize only EN locations: the APs are placed at the $N$ cluster centers; while the EN placement is optimized based on the AP locations using Algorithm \ref{41}.
  \item Local searching algorithm (LS) method introduced in Section IV, where the initial EN and AP locations are set according to the CC method and the best-performing deployment solution obtained during the searching iterations is used.
\end{itemize}

Evidently, we can see that the proposed alternating optimization has the best performance among the methods considered. Specifically, significant performance gain is observed for AltOpt over optimizing EN placement only. The LS method has relatively good performance compared to AltOpt, especially when $M$ is small, but the performance gap increases with $M$ due to the increasing probability of being trapped at local maximums with a larger $M$. The CC scheme has the worst performance as it neglects the disparity of energy harvesting/consumption rates among the WDs and precludes the case where multiple AP/ENs can be placed in a cluster. In practice, Fig. \ref{68} can be used to evaluate the deployment cost of each algorithm. For instance, when $P_r\geq -0.1$mW is required, we see that the AltOpt ($L=10$) on average needs $6$ ENs, the LS method needs $8$ ENs, optimizing EN placement only requires $9$ ENs, and the CC method needs $20$ ENs, with the same number of information APs deployed (i.e., $N=6$). The above results show that, when the ENs and APs are separated, significant performance gain can be obtained by jointly optimizing the placements of ENs and APs, especially for large-size WPCNs that need a large number of ENs and APs to be deployed. In addition, as optimizing only EN locations corresponds to a special case of Algorithm \ref{43} with $L=1$, we can see that the performance gain is significant when $L$ increases from $1$ to $10$. However, the performance improvement becomes marginal as we further increase $L$ from $10$ to $20$. In practice, good system performance can be obtained with relatively small number of alternating optimizations, e.g., $L=10$ in our case.

\begin{figure}
\centering
  \begin{center}
    \includegraphics[width=0.6\textwidth]{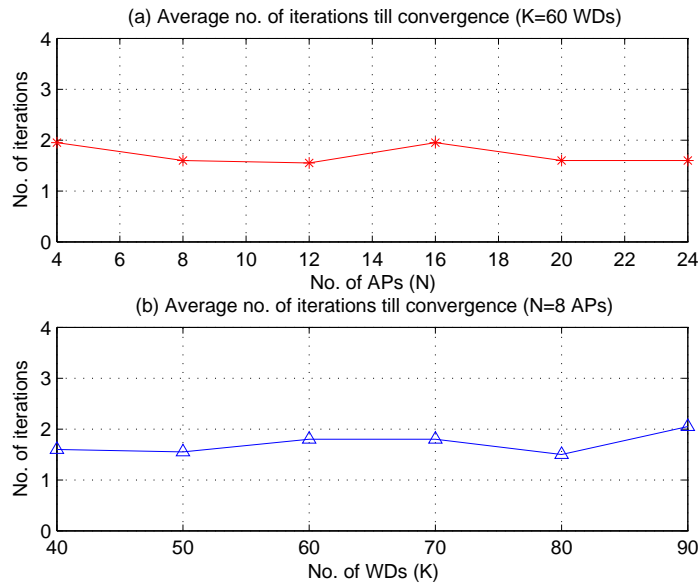}
  \end{center}
  \caption{The average number of WD-AP association assumptions made before convergence of Algorithm \ref{42}, as a function of (a) $N$ under fixed $K=60$; and (b) as a function of $K$ under fixed $N=8$.}
  \label{64}
\end{figure}

We then show in Fig.~\ref{64} the convergence rate of Algorithm \ref{42}, for which the convergence is proved in an asymptotic sense in the Appendix. In particular, we plot the average number of iterations (WD-AP association assumptions) used until the algorithm converges. Here, we investigate the convergence rate when either the number of APs ($N$) or WDs ($K$) varies. With fixed $K$ in Fig.~\ref{64}(a), we see that the number of iterations used till convergence does not vary significantly as $N$ increases. Similarly in Fig.~\ref{64}(b), with a fixed $N=8$, we do not observe significant increase of iterations when $K$ increases from $40$ to $90$. Besides, all the simulations performed in Fig.~\ref{64} use at most $7$ iterations to converge. Therefore, we can safely estimate that the number of iterations used till convergence is of constant order, i.e., $O(1)$, which leads to the complexity analysis of Algorithm \ref{43} in Section IV.C and Algorithm \ref{44} in Section V.C.

\begin{figure}
\centering
  \begin{center}
    \includegraphics[width=0.6\textwidth]{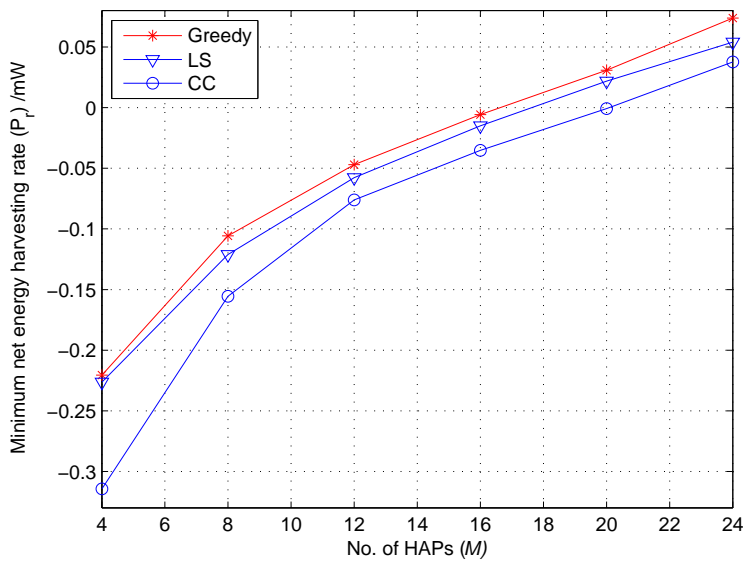}
  \end{center}
  \caption{Performance comparison of the co-located AP and EN (HAP) placement methods.}
  \label{69}
\end{figure}

\subsection{Co-located EN and AP Deployment}
Next, we evaluate in Fig.~\ref{69} the performance of the proposed Algorithm \ref{44} for co-located ENs and APs, where the value of $P_r$ achieved by Algorithm \ref{44} is plotted against the number of HAPs used ($M$). In particular, we compare its performance with that of LS (with $M$ cluster centers as the initial searching points) and the CC placement method, i.e., the HAPs are placed at the $M$ cluster centers. We can see that the proposed greedy algorithm in Algorithm \ref{44} has the best performance among the methods considered. Nonetheless, the performance gaps over the LS and CC methods are relatively small compared to that in Fig.~\ref{68}. An intuitive explanation is that the doubly-near-far phenomenon for co-located EN and AP renders the optimal HAPs placement to be around the cluster centers. By comparing Fig.~\ref{68} and Fig.~\ref{69}, we can see the evident performance advantage of using separated ENs and APs over HAPs. For instance, the $P_r$ achieved by $6$ ENs and $6$ APs in Fig.~\ref{68} is $-0.1$mW, while that achieved by $6$ HAPs (equivalent to $6$ ENs and $6$ APs being co-located) is only around $-0.17$mW.

\begin{figure}
\centering
  \begin{center}
    \includegraphics[width=0.6\textwidth]{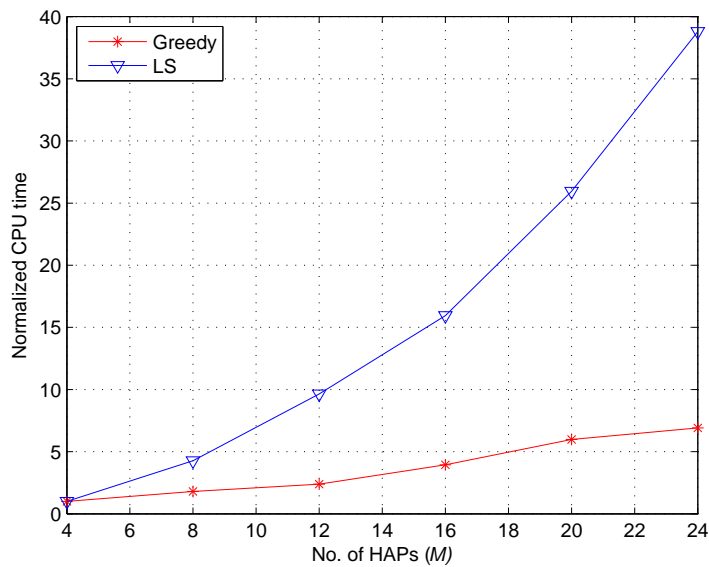}
  \end{center}
  \caption{Comparison of CPU time of the LS and Greedy methods for HAP placement optimization.}
  \label{70}
\end{figure}

Although the greedy algorithm and the LS method perform closely, they differ significantly in the computational complexity. To better visualize the growth rate of complexity, we plot in Fig.~\ref{70} the normalized CPU time of the two methods, where each point on the figure is the normalized against the CPU time of the respective method when $M=4$. Clearly, we can see that the complexity of the greedy algorithm increases almost linearly with $M$, where the CPU time increases approximately $6$ times when $M$ increases from $4$ to $24$. This matches our complexity analysis for Algorithm \ref{44} in Section V.C that the complexity increases almost linearly in $M$ when $K$ is much larger. The LS method, however, has a much faster increase in complexity with $M$, where the CPU time increases by around $39$ times when $M$ increases from $4$ to $24$. Therefore, even in a large-size WPCN with large $M$, the computation time of the proposed greedy algorithm is still moderate, while this may be extremely high for the LS method, e.g., couple of minutes versus several hours for $M =50$.

\subsection{Case Study: Separated Versus Co-located EN and AP}
Finally, we present a case study to compare the cost of node placement achieved by using either separated or co-located ENs and APs. Here, we consider a WPCN with $60$ WDs uniformly placed within the $24 m \times 24 m$ box region, where the detailed placement is omitted due to the page limit. For the case of separated ENs and APs, we use Algorithm \ref{43} to enumerate $(M,N)$ pairs that can satisfy a given net energy harvesting performance requirement $\gamma$, and select the one with the minimum cost as the solution. For the case of HAPs, we use Algorithm \ref{44} to find the minimum $M$ that can satisfy the performance requirement $\gamma$. A point to notice is that, the obtained deployment solutions are sub-optimal to the min-cost deployment problems with either separated or co-located ENs and APs, i.e., the cost of the optimal deployment solution can be lower, because Algorithms \ref{43} and \ref{44} are sub-optimal to solve problems (\ref{9}) and (\ref{14}), respectively. More effort to further improve the solution performance is needed for future investigations.

\begin{figure}
\centering
  \begin{center}
    \includegraphics[width=0.6\textwidth]{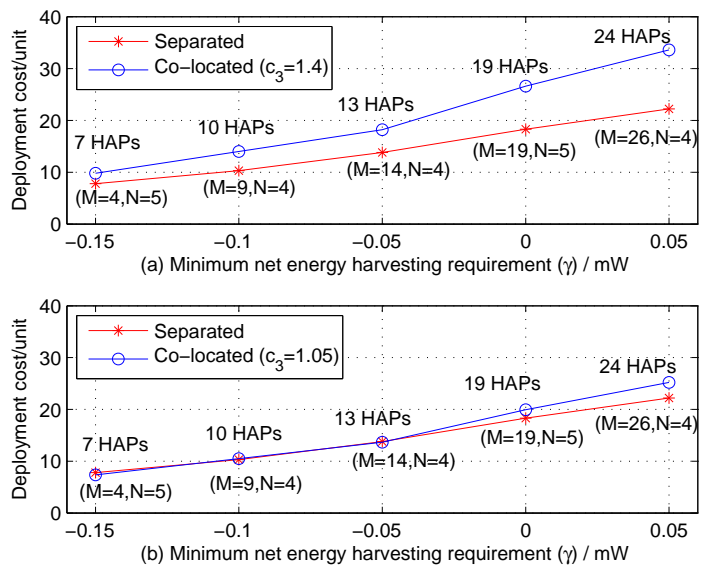}
  \end{center}
  \caption{Comparison of the minimum deployment costs achieved by separated/colocated EN and AP placement optimization methods, where in (a) $c_3=1.4$ and in (b) $c_3=1.05$. }
  \label{72}
\end{figure}

In Fig.~\ref{72}(a), we show the minimum deployment costs achieved by the two methods under different performance requirement $\gamma$. The number of nodes used by both methods are also marked in the figure. With $\left\{c_1,c_2,c_3\right\} =\left\{0.7,1,1.4\right\}$, we can see that using separated ENs and APs can achieve much lower deployment cost than co-located HAPs. In another occasion in Fig.~\ref{72}(b), the two scheme achieve similar deployment cost when the cost of a HAP is decreased from $c_3=1.4$ to $1.05$. We also observe that, by allowing the ENs and APs to be separately located, we need much less energy/information access points than they are co-located to achieve the same performance, thanks to the extra freedom in choosing both the numbers and locations of energy/information access points. For instance, when $\gamma = 0$, we need $24$ separated energy/information access points ($M=19$ and $N=5$), while $38$ co-located energy/information access points by the $19$ HAPs. However, we do not intend to claim that using separated ENs and APs is better than the co-located case. Rather, we show that a cost-effective deployment plan can be efficiently obtained using the proposed methods. In practice, the choice of using either separated or co-located ENs and APs depends on a joint consideration of the node deployment costs, the network size and the performance requirement.

\section{Conclusions and Future Work}
In this paper, we studied the node placement optimization problem in WPCN, which aims to minimize the node deployment cost while satisfying the energy harvesting and communication performance of WDs. Efficient algorithms were proposed to optimize the node placement of WPCNs with the ENs and information APs being either separated and co-located. In particular, when ENs and APs are separately located, simulation results showed that significant deployment cost can be saved by jointly optimizing the locations of ENs and APs, especially for large-size WPCNs with a large number of ENs and APs to be deployed. In the case of co-located ENs and APs (HAPs), however, the performance advantage of node placement optimization is not that significant, where we observed relatively small performance gap between optimized node placement solutions and that achieved by some simple heuristics, i.e., placing the HAPs at the cluster centers formed by the WDs. In practice, separated ENs and APs are more suitable for deploying WPCNs than co-located HAPs, because of the flexibility in choosing both the numbers and locations of ENs and APs. Nonetheless, because the optimal solution to the node placement problem has not been obtained in this paper, we may expect further improvement upon our proposed methods in the future, especially for the case of HAP placement optimization.

Finally, we conclude with some interesting future work directions for the node placement problem in WPCN. First, the models considered in this paper can be extended to more general setups. Using the uplink information transmission as example, we assumed in this paper that each WD has fixed association with a single AP. In practice, dynamic frequency allocation can be applied to enhance the spectral efficiency, where a WD can transmit to different APs, even multiple APs simultaneously, in different transmission blocks. Besides, instead of assuming each AP can serve infinite number of WDs, we can allow each AP to serve finite number of WDs. In addition, we may also consider the presence of uplink co-channel interference due to frequency reuse in WPCN. The extensions require adding corresponding constraints or changing the expression of energy consumption model in the problem formulation of this paper. Second, it is interesting to consider the hybrid node deployment problem that uses both co-located and separated ENs/APs. Third, it is practically important to consider the node placement problem with location constraints, e.g., some areas that may forbid the ENs/APs to be placed. In addition, the density of ENs may be constrained to satisfy certain safety consideration on RF power radiation.

\appendix[Proof of the Convergence of Algorithm \ref{42}] \label{59}
Let $\left\{t^{(l)},\mathbf{v}_j^{(l)},j=1,\cdots,N\right\}$ denote the optimal solutions of (\ref{27}) calculated from the $l$-th ($l\geq 1$) set of assumptions made on the WD-AP associations, denoted by $j_k^{(l)}$, $k=1\cdots,K$. Let $\mathcal{K}^{(l)}$ denote the set of WDs to which the optimal solution $\mathbf{v}_j^{(l)}$'s contradict with the WD-AP assumptions (we consider only $\mathcal{K}^{(l)}\neq \emptyset$, because otherwise the algorithm has reached its optimum), i.e.,
\begin{equation}
\small
\left\{\mathcal{K}^{(l)}: k=1\cdots,K, \ ||\mathbf{v}_{j_k^{(l)}}-\mathbf{w}_k|| > \min_{j=1,\cdots,N} ||\mathbf{v}_{j}^{(l)}- \mathbf{w}_k|| \right\}.
\end{equation}
According to the proposed trial-and-error method, $j_k^{(l+1)}$ is set for each $k=1,\cdots,K$ as
\begin{equation}
\label{29}
j_k^{(l+1)} =
\begin{cases}
j_k^{(l)}, & k\notin \mathcal{K}^{(l)},\\
\arg \min_{j=1,\cdots,N} ||\mathbf{v}_{j}^{(l)}- \mathbf{w}_k||, & k \in \mathcal{K}^{(l)}.
\end{cases}
\end{equation}

Let $\hat{t}^{(l+1)}$ denote the minimum net energy harvesting rate among all the WDs given the updated WD-AP association $j_k^{(l+1)}$'s and the current AP locations $\mathbf{v}_j^{(l)}$'s, i.e.,
\begin{equation}
\hat{t}^{(l+1)} = \min_{k=1\cdot,K} \left(\lambda_k - a_{1,k}- a_{2,k} ||\mathbf{v}_{j_k^{(l+1)}} - \mathbf{w}_k||^{d_U}\right).
\end{equation}
We can see that $\hat{t}^{(l+1)} \geq t^{(l)}$ because the update of WD-AP associations in (\ref{29}) does not increase the energy consumption rate of any WD achived by assuming $j_k^{(l)}$, $k=1\cdots,K$. Besides, $\left\{\hat{t}^{(l+1)}, \mathbf{v}_j^{(l)},j=1,\cdots,N\right\}$ is a feasible solution of (\ref{27}) under the association assumption $j_k^{(l+1)}$'s. Therefore, the optimal solution $\left\{t^{(l+1)},\mathbf{v}_j^{(l+1)},j=1,\cdots,N\right\}$ calculated from the association assumption $j_k^{(l+1)}$'s will lead to $t^{(l+1)}\geq  \hat{t}^{(l+1)} \geq t^{(l)}$. In other words, the optimal objective of (\ref{27}) is non-decreasing in each trial-and-error update of WD-AP associations. This, together with the fact that the optimal value of (\ref{27}) is bounded, leads to the conclusion that the proposed trial-and-error method is convergent.

\end{document}